\documentclass[10pt,journal]{IEEEtran}
\usepackage{color}
\usepackage{epsfig}
\usepackage{amsmath, amsthm}
\usepackage{mathtools}  
\mathtoolsset{showonlyrefs}
\usepackage{amssymb}
\usepackage{amscd}
\usepackage{threeparttable}
\usepackage{graphicx}
\usepackage{subfigure}

\providecommand{\norm}[1]{\lVert#1\rVert}

\begin{document}
\title{Adaptive Control for Unknown Heterogeneous Vehicles Synchronization with Unstructured Uncertainty}
\author{Miguel F. Arevalo-Castiblanco, D. Tellez-Castro, J. Sofrony and Eduardo Mojica-Nava}

\maketitle

\begin{abstract}
The cooperative control applied to vehicles allows the optimization of traffic on the roads. There are many aspects to consider in the case of the operation of autonomous vehicles on highways since there are different external parameters that can be involved in the analysis of a network. In this paper, we present the design and simulation of adaptive control for a platoon with heterogeneous vehicles, taking into account that not all vehicles can communicate their control input, and in turn include structured nonlinear uncertainty input parameters.
\end{abstract}

\section{Introduction}
The study of autonomous vehicles has increased in conjunction with automotive production over the last few years. For the analysis of vehicles and traffic, cooperative control is a tool widely worked lately, Cooperative Adaptive Cruise Control (CACC) is a vehicle control methodology that shares information of vehicle positions and velocities with its neighbors through a sensor system coupled \cite{Ploeg2014}. For a correct operation of cooperative control, it is necessary to consider external aspects such as the loss of communication among vehicles and its modeling as heterogeneous agents with dynamics that may be unknown \cite{wren}. 
Different control strategies have been developed for vehicle platoons to guarantee the behavior synchronization, starting its developments in the 60s with \cite{Levine1966}. One of the most common strategies for this type of systems is to include a reference model that is replicated by the vehicles in the network, Model Reference  Adaptive Control (MRAC) is a control strategy commonly used at state feedback for an agent to be adapted to a reference dynamics \cite{Baldi20182} \cite{Ioannou1996}. This strategy is classified as direct or indirect control, where the direct control adjusts the controller with all known parameters, and the indirect one in which an estimation is made based on the lack of knowledge of the agent's dynamics \cite{Nguyen2018L}. The extension of this theory to the distributed level allows synchronizing a network with heterogeneous and unknown dynamics using only local information from the neighbors \cite{Frasca2018}.

Similarly, it is important to consider the presence of uncertainty parameters such as input disturbance, in practice, vehicles networks usually present different uncertainties that make the application of the designed controllers difficult. The design of a controller that counteracts this uncertainty will strengthen its operation \cite{Nguyen2008}. The most common investigations carried out so far work with bounded disturbances that can be suppressed under optimal or robust methodologies \cite{Ioannou1996} \cite{Nguyen2018}. However, in practice, there are few cases where this type of uncertainty is present. Neural networks approximation allows estimating an structured uncertainty so that it is canceled from a linear parameterized term included in the control law \cite{Nguyen2018L}.

Another characteristic to take into account in this type of network is the lack of communication that may occur between the agents, the physical faults or limitations of the sensors on board, hinder the correct communication of all the agents in the network \cite{Zheng2019}. The design of a controller that estimates the neighbors input to be communicated between agents allows an adequate operation of the controller even when there are communication failures. Some developments have focused on generating an average virtual agent for those cases. However, with the evolution of the dynamics in adaptive control, this can lead to errors in agent's synchronization \cite{Tao2003}. The use of an input estimator allows correcting this lack of communication with the inclusion of a new adaptive law that evolves according to a predefined agents dynamic \cite{Frasca2018}.

Some authors have worked only with MRAC with neural network approximation, or with heterogeneous agents with input estimator \cite{Hongjie2008} \cite{Frasca2018}. In this case, the controller must be able to adjust some matching conditions to replicate its dynamics with respect to the reference and its neighboring agents, even when there is no communication among them, in the same way, it must allow canceling the input uncertainties through a non-linear approach.

The main contribution of this work is the development of an adaptive controller for agents with structured nonlinear input uncertainty and input estimation. First, a distributed control law that allows eliminating nonlinear input uncertainties is developed. Second, an input estimator is added to counteract the lack of communication present in the network. Third, each controller is validated to ensure that all signals are bounded. Finally, the simulation results are presented in the context of CACC.

The rest of the paper is organized as follows, in Section II the formulation of the problem is made. Section III exposes the distributed control law with the uncertainty approximation by neural networks, in Section IV the distributed estimator input is developed. Section V shows the numerical example of the vehicle platoon in front of the established topics and finally in Section VI the conclusions and future work are presented.

\textit{Notation:} The notation used for matrices and vectors are $X$ and $x$ respectively. $X^\top$ and $x^\top$ describe the transpose of a matrix or a vector. The Euclidean norm of a signal is defined as $\norm{X}^2=\sum_{i=1}^n|{x_i}|^2$. We describe $A^{-\top}=(A^\top)^{-1}$ as the inverse of a transposed matrix. The trace of an square matrix $X$ is defined as $\text{tr}(X)$. A directed graph is defined as the pair $(\mathcal{V},\mathcal{E})$, where $\mathcal{V}$ is the nodes set of the graph, and $\mathcal{E} \in \mathcal{V}\times \mathcal{V}$ is the communication edges set. The adjacency matrix is defined as $A=[a_{ij}]$ where $a_{ii}=0$ and $a_{ij}=1$ if $(j,i) \in \mathcal{E}$, with $i \neq j$ .

\section{Problem Formulation}
To facilitate the presentation of the results, this section contextualizes the problem of synchronize heterogeneous agents with nonlinear input uncertainty and in the case where the agents do not have access to the neighbors inputs. Each agent is represented with dynamics.
\begin{equation}
\dot{x}_i=A_ix_i+b_i(u_i+f_i(x_i)), \hspace{0.5cm} i \in \left[1,\;...\;,N\right],
\label{eq1}
\end{equation}
where $x_i$ $\in$ $\mathbb{R}^n$ are the agent's states, $u_i$ $\in$ $\mathbb{R}^p$ is its input, $A_i$ is an unknown matrix related to the agent's states, $b_i$ are known vectors with possibly heterogeneous agents ($A_i{\not=}A_j$ and $b_i{\not=}b_j$ ), and $f_i\colon\,\mathbb{R}^n\to\mathbb{R}^p$ is a bounded nonlinear input uncertainty that behaves like a disturbance. $f_i(x_i)$ needs to be a Lipschitz function. The reference model is described as
\begin{equation}
\dot{x}_0={A_0}x_0+{b_0}r,
\label{eq2}
\end{equation}
where $x_0$ $\in$ $\mathbb{R}^n$ is the state, $r$ $\in$ $\mathbb{R}^p$ is the reference, and $A_0$ and $b_0$ are the matrices of the reference model. 
The following assumptions describe the matching conditions and characteristics to ensure the inclusion of a cooperative MRAC. These conditions allow associating an agent with a reference model and with its neighbors to match the dynamics and replicate the same behavior.

\textbf{Assumption 1.} The vector $k_{mi}^*$ and the scalar $k_{ri}^*$ exist and are defined as
\begin{align}
A_0&=A_i+{b_i}k_{mi}^{*\top},\\
b_0&={b_i}k^*_{ri}.
\label{eq3}
\end{align}

Constants in \eqref{eq3} are known as feedback matching conditions.

\textbf{Assumption 2.} The vector $k_{mij}^*$ and the scalar $k_{rij}^*$ exists and are defined such that
\begin{align}
A_i&=A_j+{b_j}k_{mij}^{*\top}, \\
b_i&={b_j}k_{rij}^* .
\label{eq4}
\end{align}

Constants $k^*_{mij}$ and $k^*_{rij}$ in \eqref{eq4} are known as coupling matching conditions. 

\textbf{Assumption 3.} The communication graph is acyclic and must contain at least one spanning tree where the leader is connected.

From these assumptions, the problem is defined in a concrete way.

\textbf{Problem.} Consider $N$ agents with dynamics \eqref{eq1}, a reference model \eqref{eq2}, and Assumptions 1-3 verified. So, the objective of the control is to achieve that all closed-loop signals are bounded according to $t \xrightarrow{} \infty$ for each agent, even in cases where there is no communication of the control input between agents.

\section{Adaptive Synchronization with Neural Network Approximation}
In this section, Assumptions 1-3 are taken as a basis together with a nonlinear structured uncertainty parameter to be approximated and canceled through neural networks. This neural network maintains a constant closed loop connectivity called recurrent neural network.

\textbf{Proposition 1:} 
Considering the system \eqref{eq1} where the function $f_i(x)$ is approximated by a parameterized linearly function

\begin{equation}
\theta^{*\top}_i\phi_i-\epsilon^*_i,
\end{equation}
where $\epsilon^*_i \in \mathbb{R}^n$ is defined as the ideal of the approximation error, $\theta^{*\top}_i \in \mathbb{R}^{n{\times}p}$ is the ideal of the neural network related to the adaptive law and $\phi_i\colon \mathbb{R}^n \to \mathbb{R}^p$ is a known bounded basis function obtained from neural networks. Then, Using classical model reference adaptive control methodology \cite{Tao2003} it is possible to synchronize agent 1 to a reference model by the controller 

\begin{equation}
u_1={k^\top_{m1}}x_1+k_{r1}r-\theta^{\top}_1\phi_1(W^\top_1\bar{x}_1),
\label{eq6}
\end{equation}
and the adaptive laws
\begin{align}
{\dot{k}^\top_{m}}&=-\text{sgn}\left({k_{ri}}^*\right){\gamma}\: {b^\top_0}P\left(x_1-x_0\right)x_1^\top,\\
{\dot{k}_{r}}&=-\text{sgn}\left({k_{ri}}^*\right){\gamma}\: {b^\top_0}P\left(x_1-x_0\right)r,
\end{align}
where the scalar $\gamma>0$ is the adaptive gain, and $P$ is a positive definite matrix satisfying
\begin{equation}
PA_0+{A^\top_0}P=-Q, \hspace{0.5cm}Q>0,
\label{eq8}
\end{equation}
and the neural networks adaptive laws
\begin{align}
\dot{\theta}_{1}=&-\gamma\phi_1(W^\top_i\bar{x}_1)(x_1-x_0)^{\top}Pb_1,\\
\dot{W}_1=&-\gamma\bar{x}_1(x_1-x_0)^{\top}Pb_1V^\top\sigma(W^\top_1\bar{x}_1),
\label{nn1}
\end{align}
$\theta_{1}$ and $W_1$ are weight adaptive matrices, $V \in \mathbb{R}^{m{\times}n}$ is a bias vector, $\bar{x}_1=[1\hspace{0.2cm}x^\top_1]^{\top} \in \mathbb{R}^{n+1}$, $\phi_1(W^\top_1\bar{x}_1)=[1\hspace{0.2cm}\sigma^\top_1(W^\top_1\bar{x}_1)]^\top \in \mathbb{R}^{m+1}$ with $\sigma_1(x_1)$ as a sigmoidal function described by
\begin{equation}
\sigma_1(x_1)=\frac{1}{1+e^{-ax_1}},
\end{equation}

\textit{Proof:} It follows from \cite{Nguyen2018L}.

From Proposition 1, it is possible to extend the theory in a distributed way and including agents that do not have direct communication with the leader.

\textbf{Theorem 1:} Consider $N$ agents with dynamics \eqref{eq1}, where only the agent 1 has direct communication with the reference as in Proposition 1, the other agents employ the following control law

\begin{equation}
\begin{split}
u_i&=\alpha(\sum_{j=1}^{N}a_{ij}{{k}^\top_{mij}}x_j+k_{mi}\sum_{j=1}^{N}a_{ij}(x_i-x_j)+\ldots\\
&\ldots+\sum_{j=1}^{N}a_{ij}k_{rij}u_j-\theta^{\top}_i\phi_i(W^\top_i\bar{x}_i)),
\end{split}
\label{eq5}
\end{equation}
with $\alpha=\frac{1}{\sum_{j=1}^{N}a_{ij}}$ and the MRAC adaptive laws
\begin{align}
{\dot{k}^\top_{mij}}=&-\text{sgn}({k^*_{ri}})\gamma\: {b^\top_0} P\left[\sum_{j=1}^{N}a_{ij}(x_i-x_j)\right]x^\top_i,\\
{\dot{k}^\top_{mi}}=&-\text{sgn}({k^*_{ri}})\gamma\: {b^\top_0}P\left[\sum_{j=1}^{N}a_{ij}(x_i-x_j)\right]\ldots\\
&\hspace{2.1cm}\ldots{\left[\sum_{j=1}^{N}a_{ij}(x_i-x_j)\right]}^\top,\\
\dot{k}_{rij}=&-\text{sgn}({k^*_{ri}})\gamma\: {b^\top_0}P\left[\sum_{j=1}^{N}a_{ij}(x_i-x_j)\right]u_i.
\label{eqal}
\end{align}

And the neural networks adaptive laws
\begin{align}
\dot{\theta}_{i}=&-\gamma\phi_i(W^\top_i\bar{x}_i)(x_i-x_j)^{\top}Pb_i,\\
\dot{W}_i=&-\gamma\bar{x}_i(x_i-x_j)^{\top}Pb_iV^\top\sigma(W^\top_i\bar{x}_i),
\label{eq13}
\end{align}
with $\sigma_i(x_i)$ as a sigmoidal function described by
\begin{equation}
\sigma_i(x_i)=\frac{1}{1+e^{-ax_i}}.
\end{equation}

then, the control law \eqref{eq5} guarantees that all synchronization errors are bounded.

\textit{Proof:} The main idea of this proof is to validate that the convergence error of an agent that has an structured nonlinear uncertainty is bounded. For this, the error is defined as $e_{ij}=x_i-x_j$ and its dynamics is

\begin{align}
\dot{e}_{ij}&=A_me_{ij}b_i[u_i-k^{*\top}_{mij}x_j-k^{*\top}_{mi}e_{ij}-k^{*\top}_{rij}u_j-\theta^*_i\phi_i-\ldots\\
&\ldots-\epsilon^*_i+\theta^*_j+\epsilon_j],
\label{e_d}
\end{align}
taking the matching conditions of the Assumptions 1-2 and with $\tilde{k}_{mij}=k_{mij}-k^*_{mij}$; $\tilde{k}_{mi}=k_{mi}-k^*_{mi}$;
$\tilde{k}_{ri}=k_{ri}-k^*_{ri}$; $\tilde{k}_{rij}=k_{rij}-k^*_{rij}$; $\tilde{\theta}_i=\theta_i-\theta^*_i$, and taking the following Lyapunov equation

\begin{align}
&V(e_{ij}, \tilde{k}_{mi}, \tilde{k}_{rij}, \tilde{k}_{mij}, \tilde{\theta}_i)= \sum_{i=1}^{N}\left [\sum_{j=0}^{N} a_{ij} e_{ij} \right ]^\top P\ldots\\
&\ldots\left [\sum_{j=0}^{N}  a_{ij} e_{ij} \right ]+\sum_{j=1}^{N} \text{tr}\left ( \frac{{\tilde{k}_{mi}}^\top \tilde{k}_{mi}}{\gamma\, \left | {k^*_{ri}} \right | } \right )+\ldots\\
&\ldots+\sum_{i=1}^{N} \sum_{j=1}^{N}a_{ij}\; \text{tr}\left ( \frac{{\tilde{k}_{mij}}^\top \tilde{k}_{mij}}{\gamma\, \left | {k^*_{ri}} \right | } \right )+\sum_{i=1}^{N}\sum_{j=1}^{N} a_{ij}\frac{{\tilde{k}_{ri}}^2}{\gamma\, \left | {k^*_r} \right | }+\ldots\\
&\ldots+\text{tr}(\tilde{\theta}^\top_i\gamma ^{-1}\tilde{\theta}_i),
\label{eqlyap}
\end{align}
where $j=0$ is used as a representation of the reference. The derivative of \eqref{eqlyap} along \eqref{e_d} can be obtained as
{\small\begin{align}
	\dot{V}&=\sum_{i=1}^N\left[\sum_{j=0}^N{a_{ij}e_{ij}}\right]^\top(PA_0+A^\top_0P)\left[\sum_{j=0}^N{a_{ij}e_{ij}}\right]+\ldots\\
	&\ldots+2\left[\sum_{j=0}^N{a_{ij}e_{ij}}\right]^{\top}Pb_i\ldots\\
	&\ldots\left[\sum_{i=1}^N{a_{ij}{\tilde{k}_{mij}}^{\top}x_i}+{\tilde{k}_{mi}}^\top\sum_{i=1}^N{a_{ij}e_{ij}}+\sum_{i=1}^N{a_{ij}\tilde{k}_{rij}u_i}-\theta^\top_i\phi_i+\epsilon^*_i\right]+\ldots\\
	&\ldots+\sum_{i=1}^N{\text{tr}\left(\frac{{\tilde{k}_{mi}}^\top\gamma^{-1}\dot{\tilde{k}}_{mi}}{|k^*_{ri}|}\right)}+\sum_{i=1}^N{\text{tr}\left(\frac{{\tilde{k}_{mij}}^\top\gamma^{-1}\dot{\tilde{k}}_{ij}}{|k^*_{ri}|}\right)}+\ldots\\
	&\ldots+\sum_{i=1}^N{\sum_{j=1}^N{a_{ij}\frac{\tilde{k}_{rij}\gamma^{-1}\dot{\tilde{k}}_{rij}}{|k^*_{ri}|}}}-\ldots\\
	&\ldots-2\sum_{i=1}^N\sum_{j=1}^N\text{tr}\left({\tilde{\theta}_i}^\top\phi_ie^\top_{ij}Pb_i\right),
	\label{dlyapg}
	\end{align}}
reducing \eqref{dlyapg} we have
\begin{align}
\dot{V}&=-\sum_{i=1}^N\left[\sum_{j=0}^N{a_{ij}e_{ij}}\right]^{\top}Q\left[\sum_{j=0}^N{a_{ij}e_{ij}}\right]+\ldots\\
&\ldots+2\left[\sum_{j=0}^N{a_{ij}e_{ij}}\right]^{\top}Pb_i\left({\tilde{\theta}_i}^\top\phi_i+\epsilon^*_i\right),
\end{align}
then
\begin{align}
\dot{V}&=-\sum_{i=1}^N\left[\sum_{j=0}^N{a_{ij}e_{ij}}\right]^{\top}Q\left[\sum_{j=0}^N{a_{ij}e_{ij}}\right]+\ldots\\
&\ldots+2\left[\sum_{j=0}^N{a_{ij}e_{ij}}\right]^{\top}Pb_i\epsilon^*_i\leq-\sum_{i=1}^N\lambda_{\min}\left(Q\right)\ldots\\
&\ldots\sum_{j=1}^N\norm{e_{ij}}^2+2\sum_{i=1}^N\sum_{j=1}^N\norm{Pb_i}\norm{e_{ij}}\epsilon^*_0.
\end{align}

On the case $\dot{V}\leq0$ if
\begin{align}
-&\sum_{i=1}^N\lambda_{\min}\left(Q\right)\sum_{j=1}^N\norm{e_{ij}}^2+2\sum_{i=1}^N\sum_{j=1}^N\norm{Pb_i}\norm{e_{ij}}\epsilon^*_0\leq0\Rightarrow\ldots\\
&\ldots\Rightarrow\sum_{i=1}^N\sum_{j=1}^N\norm{e_{ij}}\geq\frac{2\norm{Pb_i}\epsilon^*_0}{\lambda_{\min}Q},
\end{align}
so the conditions are then met to ensure that closed loop synchronization error of an agent with structured nonlinear uncertainty is bounded. $\blacksquare$

\section{Adaptive Synchronization with unknown estimated input}

In this section we analyze the case where an agent $i$ has no communication from the input $u_j$ of its neighboring agents. The control law in this case is defined as

\begin{align}
u_i&=\alpha(\sum_{j=1}^{N}a_{ij}{{k}^\top_{ij}}x_j+k_{mi}\sum_{j=1}^{N}a_{ij}(x_i-x_j)+\ldots\\
&\ldots+\sum_{j=1}^{N}a_{ij}\hat{u}_{ji}-\theta^{\top}_i\phi_i),
\label{eq12}
\end{align}
with the adaptive laws \eqref{eqal}, \eqref{eq13}, where $\hat{u}_j$ represents the agent input $u_j$, this estimation allows to suppress the calculation of $k_{rij}$ that relates the entries between neighboring agents. The dynamics of the input estimation $u_i$ is determined by
\begin{equation}
\dot{\hat{u}}_{ji}=-\text{sgn}(k_{ri}^*)\gamma{b'}_0P\left[\sum_{i=1}^{N}a_{ij}(x_i-x_j)\right].
\label{eq16}
\end{equation}

\textbf{Remark 1:} The estimator for an agent $j$ is calculated for each neighbor connected to the agent $i$ in a distributed way.

\textbf{Theorem 2:} Consider a network of heterogeneous vehicles with unknown dynamics \eqref{eq1} and a reference model \eqref{eq2} with constant reference signal, controller \eqref{eq12} and adaptive laws \eqref{eqal} and \eqref{eq13}, then, all closed loop signals are bounded.

\textit{Proof:} The proof is performed to validate that the synchronization error of an agent that does not have communication with its neighbors is bounded. For this, the error dynamics $e_{ij}$ is defined as
\begin{align}
\dot{e}_{ij}&=A_me_{ij}b_i[u_i-k^{*\top}_{mij}x_j-k^{*\top}_{mi}e_{ij}-u^*_{ji}-\theta^*_i\phi_i-\epsilon^*_i+\ldots \\
&\ldots+\theta^*_j+\epsilon_j],
\label{e_d2}
\end{align}, with $\tilde{u}_{ji}=u_{ji}-u^*_{ji}$, and taking the following Lyapunov equation

\begin{align}
V(&e_{ij}, \tilde{k}_{mi}, \tilde{k}_{rij}, \tilde{k}_{mij}, \tilde{\theta}_i)= \sum_{i=1}^{N}\left [\sum_{j=0}^{N} a_{ij} e_{ij} \right ]^\top P\ldots\\
&\ldots\left [\sum_{j=0}^{N}  a_{ij} e_{ij} \right ]+\sum_{j=1}^{N} \text{tr}\left ( \frac{{\tilde{k}_{mi}}^\top \tilde{k}_{mi}}{\gamma\, \left | {k^*_{ri}} \right | } \right )+\ldots\\
&\ldots+\sum_{i=1}^{N} \sum_{j=1}^{N}a_{ij}\; \text{tr}\left ( \frac{{\tilde{k}_{mij}}^\top \tilde{k}_{mij}}{\gamma\, \left | {k^*_{ri}} \right | } \right )+\sum_{i=1}^{N}\sum_{j=1}^{N} a_{ij}\frac{{\tilde{u}_{ji}}^2}{\gamma\, \left | {k^*_r} \right | }+\ldots\\
&\ldots+\text{tr}(\tilde{\theta}^\top_i\gamma ^{-1}\tilde{\theta}_i),
\label{eqlyap2}
\end{align}
where $j=0$ is used as a representation of the reference. The derivative of \eqref{eqlyap2} along \eqref{e_d} is
{\small\begin{align}
	\dot{V}&=\sum_{i=1}^N\left[\sum_{j=0}^N{a_{ij}e_{ij}}\right]^\top(PA_0+A^\top_0P)\left[\sum_{j=0}^N{a_{ij}e_{ij}}\right]+\ldots\\
	&\ldots+2\left[\sum_{j=0}^N{a_{ij}e_{ij}}\right]^{\top}Pb_i\ldots\\
	&\ldots\left[\sum_{i=1}^N{a_{ij}{\tilde{k}_{mij}}^{\top}x_i}+{\tilde{k}_{mi}}^\top\sum_{i=1}^N{a_{ij}e_{ij}}+\sum_{i=1}^N{a_{ij}\tilde{u}_{ji}}-\theta^\top_i\phi_i+\epsilon^*_i\right]+\ldots\\
	&\ldots+\sum_{i=1}^N{\text{tr}\left(\frac{{\tilde{k}_{mi}}^\top\gamma^{-1}\dot{\tilde{k}}_{mi}}{|k^*_{ri}|}\right)}+\sum_{i=1}^N{\text{tr}\left(\frac{{\tilde{k}_{mij}}^\top\gamma^{-1}\dot{\tilde{k}}_{ij}}{|k^*_{ri}|}\right)}+\ldots\\
	&\ldots+\sum_{i=1}^N{\sum_{j=1}^N{a_{ij}\frac{\tilde{u}_{ji}\gamma^{-1}\dot{\tilde{u}}_{ji}}{|k^*_{ri}|}}}-\ldots\\
	&\ldots-2\sum_{i=1}^N\sum_{j=1}^N\text{tr}\left({\tilde{\theta}_i}^\top\phi_ie^\top_{ij}Pb_i\right),
	\label{dlyapg2}
	\end{align}}
reducing
\begin{align}
\dot{V}&=-\sum_{i=1}^N\left[\sum_{j=0}^N{a_{ij}e_{ij}}\right]^{\top}Q\left[\sum_{j=0}^N{a_{ij}e_{ij}}\right]+\ldots\\
&\ldots+2\left[\sum_{j=0}^N{a_{ij}e_{ij}}\right]^{\top}Pb_i\left({\tilde{\theta}_i}^\top\phi_i+\epsilon^*_i\right),
\end{align}
then
\begin{align}
\dot{V}&=-\sum_{i=1}^N\left[\sum_{j=0}^N{a_{ij}e_{ij}}\right]^{\top}Q\left[\sum_{j=0}^N{a_{ij}e_{ij}}\right]+\ldots\\
&\ldots+2\left[\sum_{j=0}^N{a_{ij}e_{ij}}\right]^{\top}Pb_i\epsilon^*_i\leq-\sum_{i=1}^N\lambda_{\min}\left(Q\right)\ldots\\
&\ldots\sum_{j=1}^N\norm{e_{ij}}^2+2\sum_{i=1}^N\sum_{j=1}^N\norm{Pb_i}\norm{e_{ij}}\epsilon^*_0.
\end{align}

On the case $\dot{V}\leq0$ if
\begin{align}
-&\sum_{i=1}^N\lambda_{\min}\left(Q\right)\sum_{j=1}^N\norm{e_{ij}}^2+2\sum_{i=1}^N\sum_{j=1}^N\norm{Pb_i}\norm{e_{ij}}\epsilon^*_0\leq0\Rightarrow\ldots\\
&\ldots\Rightarrow\sum_{i=1}^N\sum_{j=1}^N\norm{e_{ij}}\geq\frac{2\norm{Pb_i}\epsilon^*_0}{\lambda_{\min}Q},
\end{align}
so we can proof that all closed loop signals of the agent described are bounded. $\blacksquare$

\section{Numerical Example}
Introducing the field of application of the proposed algorithms, the problem of a network of autonomous vehicles is raised, where each vehicle must follow the same speed pattern and maintain a distance between each one. The most well-known technology for this problem is the CACC, an extension of the Adaptive Cruise Control (ACC), where the problem of vehicles in platoon with the presence of on-board sensors arises. Each agent is modeled as a linear second order system such as

\begin{eqnarray}
\dot{x}_i = \left[ \begin{array}{cc}
0 & 1 \\
a_{1i} & a_{2i} \end{array} \right]\
x_i+b_{i} \left(\left[ \begin{array}{c}
0 \\
u_{i} \end{array} \right] + f_i(x_i)\right).
\label{eq17}
\end{eqnarray}

Where $a_{1i}$ and $a_{2i}$ are parameters of the transmission and $b_{1i}$ is a parameter of engine efficiency. These parameters are different for each vehicle, so it is considered heterogeneous agents. The input $u_i$ is the acceleration, or the force multiplied by the mass of the vehicle. The leading vehicle or reference model poses an acceleration profile that all agents must follow with a fixed distance between each one, in terms of synchronization $x_i-x_j\xrightarrow{}0$.

\begin{figure}[ht]
	\centering
	\includegraphics[width = 0.4\textwidth]{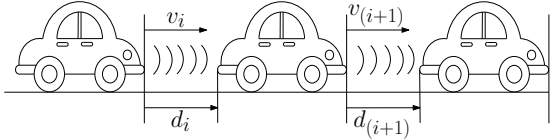}
	\caption{Vehicles platoon.}	
	\label{fig:carros}
\end{figure}

For a clearer representation, considering a platoon of $N$ vehicles as Fig. \ref{fig:carros}, where $v_i$ is the speed of the agents and $d_i$ is the distance between each vehicle. It is important to highlight four aspects of this methodology: the dynamics of the vehicles, the distributed controller, the information communicated through the network and its topology \cite{Baldi2018Ifac}.

\begin{figure}[ht]
	\centering
	\includegraphics[width = 0.2\textwidth]{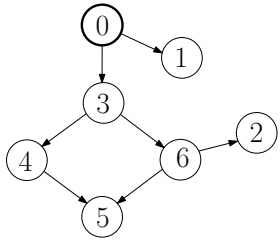}
	\caption{Leader-follower communication graph.}	
	\label{fig:1}
\end{figure}

To validate the control law, a numerical simulation is performed. Fig. \ref{fig:1} shows the digraph considered for the simulation, where the agent $0$ acts as reference model.

\begin{table}[b]
	\vspace{0.3cm}
	\centering
	\begin{tabular}{ccccc}
		\hline
		& $a_1$ & $a_2$ & $b_1$ & $x_0$  \\ 
		\hline
		$A_0$ & -0.25 & -0.5 & 1 & $[1 \hspace{0.1cm} -1]^\top$ \\  
		$A_1$ & -1.25 & 1 & 0.5 & $[1 \hspace{0.1cm} 0]^\top$ \\ 
		$A_2$ & -0.5 & 2.5 & 0.75 & $[-1 \hspace{0.1cm} 0.5]^\top$ \\ 
		$A_3$ & -0.75 & 2 & 1.5 & $[1 \hspace{0.1cm}  0]^\top$ \\ 
		$A_4$ & -1.5 & 2.5 & 1 & $[-1 \hspace{0.1cm}  1]^\top$ \\ 
		$A_5$ & -1 & 2 & 1 & $[-0.5 \hspace{0.1cm}  1]^\top$ \\ 
		$A_6$ & -0.75 & 1 & 0.5 & $[0 \hspace{0.1cm}  -1]^\top$ \\ \hline
	\end{tabular}
	\caption{Agent's Coefficients and Initial Conditions}
	\label{tab:my_label}
\end{table}

\begin{figure}[ht]
	\centering
	\includegraphics[width = 0.48\textwidth]{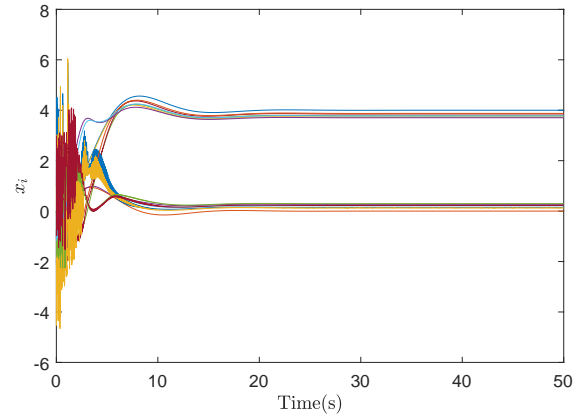}
	\caption{Agents synchronization with neural network nonlinear uncertainty approximation.}	
	\label{fig:2}
\end{figure}

\begin{figure}[htbp]
	\centering
	\includegraphics[width = 0.48\textwidth]{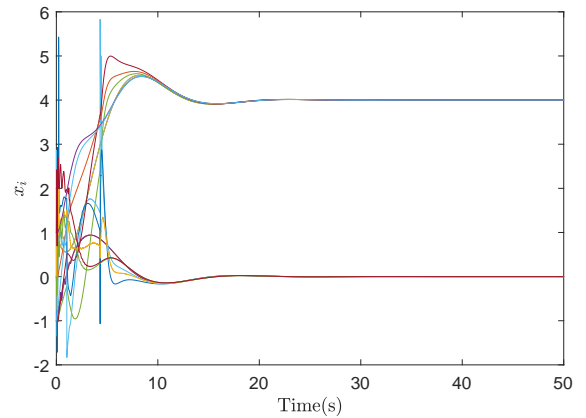}
	\caption{Agents synchronization with input estimation and neural network nonlinear uncertainty approximation.}	
	\label{fig:3}
\end{figure}

The simulation parameters used are shown in Table \ref{tab:my_label}, noting that these parameters are unknown and are used only for simulation, not for control design. 
All agents are unstable in open loop, except the reference model. For simulation purposes, the following additional parameters are necessary: $\gamma = 10$, $Q=\text{diag}(100,1)$. The matching conditions gains to the neighbors and to the reference are initialized in 0, while the gains associated to the neural network are initialized in a random value within the set $[-0.3,0.3]$. Two simulations are carried out to validate the proposed theory, a first simulation shows a network synchronization, where its agents communicate the input value between them. A second simulation shows the case where the input of the neighbors is estimated. In both cases, the followers agents have a non-linear uncertainty at the input. Fig. \ref{fig:2} shows the result of the first simulation, where the agents communicate its input between neighbors, the convergence of the states to the reference agent is observed. Fig. \ref{fig:3} shows the result of agent synchronization estimating the input of the neighbors, an asymptotic convergence to the reference model is guaranteed, with a slight increase in the oscillation in the initial seconds of simulation. It is important to highlight the presence of an overshoot in some of the followers agents, unlike conventional MRAC, derived from the initial conditions of $\theta_i$ and $W_i$ arbitrarily chosen.

\section{Conclusions and Future Work}
This work presents the development methodology of an adaptive controller for systems with structured uncertainty approximated through neural networks and with input estimator for the synchronization of heterogeneous vehicles and with partially unknown dynamics. The problem is solved based on an MRAC synchronization problem where each agent converges to the behavior of its neighbors. From the matching conditions, it is possible to replicate the dynamics of each agent according to the reference and its neighbors, even when there is no communication between them by estimating the input.
In the presence of structured uncertainty, an approximation by neural networks is developed which allows to cancel it without affecting the synchronization of each agent. An boundary analysis based on Lyapunov is performed to ensure that all closed-loop error signals are bounded. 
As future work, the extension of the theory to cyclic graphs is proposed, suppressing the switching and the loop present in the network, as well as the physical interconnection of the agents for other fields of application and the approach of the same control theory but applied as an output regulation problem.

\bibliographystyle{IEEEtran}
\bibliography{referencias}

\end{document}